\documentclass[preprint2]{aastex}

\def\ie{{\it i.e.\/}}
\def\hratio{{h_{75}^{-1}}}
\def\lhratio{{h_{75}^{-2}}}

\begin{document}

\slugcomment{Submitted to the Astronomical Journal}

\title{The X-ray Properties of Nearby Abell Clusters from the ROSAT
  All-Sky Survey: The Sample and Correlations with Optical Properties}

\author{Michael J. Ledlow} 
\affil{Gemini Observatory, Southern Operations Center, AURA, Casilla
  603, La Serena, Chile}

\author{Wolfgang Voges}

\affil{Max-Planck-Institut f{\"u}r Extraterrestrische Physik, Postfach 1312, D-85740, \\
  Garching bei M\"unchen, Germany}

\author{Frazer N. Owen}
\affil{National Radio Astronomy Observatory\altaffilmark{1}, Socorro, NM 87801}

\author{Jack O. Burns} 
\affil{Department of Astrophysical \& Planetary Sciences, University
  of Colorado, Boudler, CO 80309}

\altaffiltext{1}{The National Radio Astronomy Observatory is operated
  by Associated Universities, Inc., under contract with the National
  Science Foundation.}

\begin{abstract}
  
  We present an analysis of the X-ray emission for a complete sample
  of 288 Abell clusters spanning the redshift range $0.016 \leq z \leq
  0.09$ from the {\sl ROSAT} All-Sky Survey. This sample is based on
  our 20cm VLA survey of nearby Abell clusters.  We find an X-ray
  detection rate of 83\%.  We report cluster X-ray fluxes and
  luminosities and two different flux ratios indicative of the
  concentration and extent of the emission.  We examine correlations
  between the X-ray luminosity, Abell Richness, and Bautz-Morgan and
  Rood-Sastry cluster morphologies.  We find a strong correlation
  between $L_X$ and cluster richness coupled to a dependence on the
  optical morphological type.  These results are consistent with the
  observed scatter between X-ray luminosity and temperature and a
  large fraction of cooling flows.  For each cluster field we also
  report the positions, peak X-ray fluxes, and flux-ratios of {\bf
    all} X-ray peaks above 3$\sigma$ significance within a box of
  $2\times 2 ~\hratio$ Mpc centered on Abell's position.

\end{abstract}

\section{Introduction}

Rich clusters of galaxies are potentially powerful testbeds for
cosmological models.  They are rare, $\approx$10$\sigma$ fluctuations
in the mass spectrum and, thus, are sensitive to the underlying
cosmology.  Also, clusters are now commonly believed to have formed
recently \citep{evrard93,bryan}, so we can observe their evolution at
relatively modest redshifts.  Finally, rich clusters are luminous at
optical, X-ray, and radio wavelengths making them bright observational
targets out to $z\approx$1.

Our view of clusters of galaxies has changed dramatically over the
past decade.  We no longer see clusters as simple spherical, isolated
structures in virial and hydrostatic equilibrium.  Rather, they are
dynamic, evolving, and young systems that are strongly influenced by
their connections to the large-scale structures in which they have
formed \citep{chambers}. Their evolution is driven by the accretion
of dark matter, galaxies, and gas along filaments and via
cluster-cluster mergers \citep{bekki,burns98,kurt96,navarro,cen94}.
Evidence for the on-going accretion of subsystems, and thus relative
youth, is found in the abundant substructure observed in both optical
\citep{kolokotronis,pink96,bird} and X-ray \citep{sanders,jf99,mohr}
imaging of clusters.  Even very rich nearby clusters such as Coma,
once presumed to be relaxed and virialized, show evidence of recent
merger activity \citep{watanabe,burns94,white93}.  Extended radio
sources in clusters have also been shown to be useful probes of
cluster-cluster mergers \citep{burns98}.  Correlations between central
X-ray substructure and bent radio sources, such as wide-angle tailed
\citep{gomez}, narrow-angle tailed \citep{bliton} radio galaxies, and
cluster-wide radio haloes \citep{giovannini} reveal strong evidence of
recent mergers. These radio sources, when combined with numerical
models \citep{loken} can be used as the only currently available probe
of cluster gas dynamics.

Much of what we know about rich clusters at X-ray and radio
wavelengths has come from incomplete samples observed with a wide
variety of sensitivities and resolutions.  However, with our VLA 20-cm
survey of nearby Abell clusters \citep{led95a,owenled} and the {\it
  ROSAT} all-sky X-ray survey (RASS) \citep{voges99}, we have an
opportunity to formulate and study statistical and uniform samples of
rich clusters in the radio and the X-ray for the first time.  In
\cite{led99} we presented the X-ray luminosity function (XLF) for this
sample, and have used the observed local cluster abundance to
constrain cosmological models, at least weakly.

The usefulness of the RASS for examining the X-ray properties of rich
clusters has been previously demonstrated by several studies.
\cite{eb93} performed a cross-correlation of the RASS with the entire
ACO catalog \citep{aco} using an early release of the RASS with
uniform exposure time ($\sim$400 sec) to a flux limit of $\approx
7\times10^{-13}$ erg cm$^{-2}$s$^{-1}$ (0.1-2.4 keV).  Using a newly
developed detection algorithm (VTP), \cite{xbacs}, compiled an X-ray
flux-limited sample of the 242 X-ray Brightest Abell Clusters (XBACS)
to a flux limit of $\approx 5\times10^{-12}$ erg cm$^{-2}$ s$^{-1}$
(0.1-2.4 keV).  Briel \& Henry (1993) measured X-ray fluxes and upper
limits for a complete subsample of 145 Abell clusters with measured
redshifts.  They detected 46\% of the clusters to $4.2\times 10^{-13}$
erg cm$^{-2}$ s$^{-1}$ (0.5-2.5 keV).  \cite{bcs} compiled a
completely X-ray selected sample of the 201 brightest clusters (BCS)
with $z\leq 0.3$ and a flux-limit of $4.4\times 10^{-12}$ erg
cm$^{-2}$ s$^{-1}$ (0.1-2.4 keV). \cite{ebcs} extended the BCS (eBCS)
to a flux-limit of $2.8\times 10^{-12}$ ergs cm$^{-2}$ s$^{-1}$ and
the addition of 99 more clusters with z$<$0.3 (8 more with z$>$0.3).
Similarly, \cite{degrandi99} identified clusters in the Southern
hemisphere down to $3-4\times10^{-12}$ erg cm$^{-2}$ s$^{-1}$ (0.5-2.0
keV).  They identified 130 clusters to this limit, 101 being Abell/ACO
clusters.  The work in the southern hemisphere was followed up by
\cite{cruddace2002} and \cite{reflex} (REFLEX: The ROSAT-ESO flux-
limited X-ray Galaxy Cluster Survey).  The former concentrating on an
$\approx 1$ steradian region around the south galactic pole, the
latter making use of the 2nd reprocessing of the RASS and optical
identifications from the COSMOS digital catalogue and a followup
redshift survey. They identified 452 clusters to a limit of $3\times
10^{-12}$ erg cm$^{-2}$ s$^{-1}$ (0.1-2.4 keV).  \cite{nep} compiled a
sample of 64 clusters (nearly a third at $z>0.3$) taking advantage of
the deepest portion of the RASS at the north ecliptic pole (limit of
$f_x \sim 7.8\times 10^{-14}$ ergs cm$^{-2}$ s$^{-1}$ (0.5-2.0 keV)).
\cite{hiflugs,macs} have concentrated on the most distant (z$\geq$0.3)
and most massive clusters detected by the RASS.  See \cite{edge2003}
for a complete summary of X-ray Surveys of low-redshift clusters.

In this paper, we use the RASS to examine the X-ray properties of a
statistically complete sample of 288 northern Abell clusters with
z$\leq$0.09. Thus our sample is based on optically selected clusters,
similar to the XBACs sample but includes clusters more than 10 times
fainter (our average flux limit is $3.1\times 10^{-13}$ ergs cm$^{-2}$
s$^{-1}$ (0.5-2.0 keV)).  Despite the optical-selection, the XLF
agrees very well with X-ray selected cluster samples for $L_X \geq
10^{43}h_{50}^{-2}~\rm ergs/sec$ \citep{led99}.  The usefulness of our
sample compared to previous compilations is the abundance of
pre-existing data (our VLA 20cm survey), complete redshift
information, and its large size.  Our sample is $\approx$ 50\% larger
than the BCS within a volume that is nearly 10 times smaller in size
by virtue of the lower flux limit.

In $\S$2, we describe the construction of the Abell cluster sample
used in this study.  In $\S$3, we discuss the basic properties of the
RASS.  In $\S$4, we describe the methods used for finding and
characterizing the X-ray emission, descriminating between extended and
point-source emission, and measuring X-ray luminosities.  In $\S$5, we
summarize the results with two extended tables.  Table 1 includes
detection statistics for the clusters, while Table 2 lists {\bf all}
significant X-ray peaks found in the fields.  Also in $\S$5, we look
at the properties of known cooling flow clusters in our sample and
examine correlations of the X-ray properties with cluster richness and
optical morphology.  We summarize our conclusions in $\S$ 6. We have
used a Friedmann-Walker cosmology with $H_o$ =75 km/sec/Mpc and
$q_o$=0.5 throughout.

\section{Description of the Sample}

The sample studied in this paper was initially constructed as part of
a VLA 20cm survey of rich clusters \citep{zhao,owb,owg}.  A detailed
analysis of the optical and radio properties of these clusters is
presented in \cite{led95a,led95b,led96}, and \cite{owenled}. In short,
this sample includes all northern Abell clusters with measured
redshifts $\leq 0.09$. Clusters with high galactic extinction ($>0.1$
magnitudes at R-band, corresponding to $\log N_H > 20.73$ and roughly
to $|b| \lesssim 25^{\circ}$) were excluded.  In addition, clusters
which were very discrepant in their $m_{10}-z$ relationship ($\sim 20$
clusters) were also excluded \citep{led95a}.  In most cases, these
clusters are more distant than predicted by the $m_{10}-z$
relationship because of foreground galaxy contamination, which biased
Abell's estimate.  These selection criteria result in a sample of 288 
Abell clusters with confirmed redshifts $0.016 \leq z \leq 0.09$, and
forms a complete sample of nearby, optically-selected rich clusters
useful for statistical studies.  

Redshifts for each cluster were checked with NED and the most recent
compilation of \cite{sr99}(SR).  The redshifts for eight clusters
changed significantly as tabulated by SR from the time that we began
this project.  The changes in redshift from those listed in
\cite{led95a} from which this sample is based are: A912
($z_{old}$=0.08, $z_{new}$=0.0446), A971 ($z_{old}$=0.06,
$z_{new}$=0.0929), A1035 ($z_{old}$=0.0799, $z_{new}$=0.0684), A1927
($z_{old}$=0.0908, $z_{new}$=0.0947), A2004 ($z_{old}$=0.0640,
$z_{new}$=0.1365), and A2295 ($z_{old}$=0.0540, $z_{new}$=0.0823).
While exceeding our $z\leq 0.09$ cutoff, we have left A1927 (also in
BCS), A971, and A2004 in the tables as these are the first
measurements of cluster emission for all but A1927.  There has been
inconsistency in the literature for A1423. The cluster is catalogued
at z=0.0761 in SR (based on 3 measurements) and they make note that
\cite{crawford95} lists a galaxy at z=0.213 for the cluster redshift.
More recently, \cite{dahle2002} have found evidence from the color
magnitude sequence to support the higher z=0.21 estimate.  We have
also eliminated A1461 ($z_{old}$=0.0537), listed as unreliable in SR
at distance class = 5 and A2638 (z=0.0825 in SR), which is more likely
at the redshift of a central radio galaxy with z=0.216 \citep{olk95}.
Additionally, recent work by \cite{blakeslee} suggests that the X-ray
emission in A2152 is associated with a concentration of background
galaxies (z=0.13) based on a detailed velocity and lensing analysis.
Finally, A2589 ($z=0.0414$) while part of the 288 clusters in the
statistical sample, had essentially zero exposure time in the RASS and
so no analysis was possible.

We derived the selection-function for our sample based on the criteria
listed above.  Using the {\sl HI} map of \cite{h1map}, we first
excluded all regions with $\log N_H > 20.73$ (based on a fit to E(B-V)
vs. $\log N_H$)\citep{bh82}.  Second, we excluded regions (and
clusters) of the sky judged to be statistically incomplete by
\cite{abell58}.  Third, we restricted the sky-coverage to declination
$>-27^{\circ}$ for Abell's northern catalog.  A plot in galactic
coordinates showing the survey area is given in Figure 1a.  The
solid-line is the $\delta=-27^{\circ}$ limit.  Solid dots denote
regions of the sky which meet the selection criteria.  In Figure 1b,
we show the distribution of our sample on the same coordinate grid for
comparison.  We calculate a survey area of 4.31 steradians (14,155
deg$^2$) or 34\% of the sky.  Within the volume defined by this survey
region and the redshift limits of $0.016 \leq z \leq 0.09$, we find
the number density of clusters to be constant as a function of
richness class and redshift interval suggesting that our sample is
complete and volume-limited within the limits of Abell's selection
criteria.  These results are in agreement with \cite{briel93} and
\cite{mazure} who found Abell's catalog to be complete for all
richness classes over this range in redshift.  However, we expect some
incompleteness at the faint-end of the XLF because of our optical
selection.  Our sample is not flux-limited in the X-ray, and optically
poorer clusters with $L_X$ in the range of Richness-class 0 Abell
clusters will be missed in our sample \citep{led99}.

\section{The ROSAT All-Sky Survey}

The ROSAT All-Sky Survey (RASS) was conducted mainly during the first
half year of the ROSAT mission in 1990 and 1991 \citep{trumper}. The
ROSAT mirror system \citep{asch} and the Position Sensitive
Proportional Counter (PSPC) \citep{pfeff} operating in the soft X-ray
regime (0.1-2.4 keV) provided optimal conditions for the studies of
celestial objects with low surface brightness.  In particular, due to
the unlimited field of view of the RASS and the low background of the
PSPC, the properties of nearby clusters of galaxies can be ideally
investigated.

During the survey phase, ROSAT scanned the sky in great circles whose
planes were oriented roughly perpendicular to the solar direction.
This resulted in an exposure time varying between about 400 sec and
40,000 sec at the ecliptic equator and poles, respectively.  During
the passages through the auroral zones and the South Atlantic Anomaly
the PSPC had to be switched off, leading to a decrease of exposure
over parts of the sky. In addition, some portions of the sky received
less exposure where not enough guide stars for the automatic measuring
and control system (AMCS) of the satellite were available. As a
consequence of these various facts, the exposure map is not always
homogenous; in some small regions of the sky (0.4\%) the exposure time
can be less than 50 seconds \citep{voges99}.  We find an average
exposure time of 473 seconds for our sample.  Only one cluster (A2589)
was effectively missed completely in the survey, and therefore no
estimation of a detection threshold is possible.

For studying nearby clusters of galaxies, we extracted from the RASS
all hard photons (0.5-2.0 keV) in a 2$^{\circ}$ box centered at the
optical position of the Abell cluster to generate the photon event
image. We then calculated the vignetting corrected exposure map to
construct the exposure-corrected X-ray images.

In Figures 2a,b,c we show optical/X-ray overlays for six
average-luminosity clusters from our sample ($L_X \approx \lhratio
10^{43}~ergs~sec^{-1}$).  The X-ray images were smoothed to
$250\hratio~kpc$ resolution at the redshift of each cluster.  The
lowest contour shown is twice the estimated X-ray background (see next
section).  The optical images were obtained from the Digitized Sky
Survey (POSS I).  The field-of-view of each plot is $\approx
2.5\hratio~Mpc$.  Overlays for the entire sample are available over
the WWW at \url{http://put.the.rightaddress.here}.

\section{Image Analysis} 

\subsection{Methods for Finding the X-ray Peaks} 

In order to initially assess the quality and properties of the RASS
data for this sample, we examined each cluster individually within the
NRAO Astronomical Image Processing System (AIPS).  First, we convolved
the images to an effective spatial resolution of 250 and 500$\hratio$
kpc at the redshift of the cluster using a Gaussian smoothing function
to facilitate the identification of the X-ray peaks.  We then visually
identified peaks within a 2 Mpc diameter box centered on Abell's
optical center position.  For each peak, a surface brightness profile
was measured using the AIPS task {\sl IRING}.  The procedure was to
measure the mean and cumulative count rates in 32 equally-spaced
circular annuli out to 1$\hratio$ Mpc in radius centered on the
identified X-ray peaks (the annuli were separated by 31.25$\hratio$
kpc in radius, $\approx$2 pixels at the mean redshift of the sample).
The surface brightness profiles were measured from minimally-smoothed
($25\arcsec$ FWHM) exposure-corrected images.  From the
surface-brightness profiles, some idea of the extent and brightness of
the identified peaks could be determined.

We used these visually identified X-ray peaks only as a first-pass to
get a {\sl feel} for the data. In order to produce a more statistical
catalog of X-ray identifications within the cluster fields, we adopted
the peak-finding program {\sl SAD} (Search and Destroy) in the AIPS
package.  The {\sl SAD} program finds all sources within
some subimage whose peak is brighter than a given level.  We specified
the minimum level to be $3\sigma$ above the error in the background
(see below). Once all the possible peaks are identified, {\sl SAD}
uses a least-squares method to fit an elliptical Gaussian to each
peak.  We used these peak locations for the coordinate positions of
the identifications. We ran SAD in an automated mode on all 250 and
500$\hratio$ kpc smoothed images.  Given the (X,Y) centroids of the
significant peaks, we then processed the entire data set (\ie measured
surface-brightness profiles) in an automated batch-mode using a
custom-written AIPS procedure.  Additional stand-alone C and FORTRAN
programs were then used to extract and analyze the intensity profiles
and to calculate the X-ray luminosities.  We only used the peaks
identified from the 250$\hratio$ kpc smoothed images in the final
analysis.

We checked the identifications from SAD with our visual ones, and
computed surface-brightness profiles for any which appeared to have
been missed.  From the cumulative intensity profile we then checked
these peaks against our S/N $\geq$ 3 criteria over the $500\hratio$
kpc aperture. Only a very few ($<$5\%) of these additional
visual-selected peaks survived, mostly being confused by a low peak
imbedded in more diffuse emission and a few cases of overestimation in
the background level.  We thus judge our completeness level to be
$>$95\% for X-ray peaks in the cluster fields.  For detecting the
extended cluster emission we believe that our method is 100\% complete
within the sensitivity limits of the RASS.

\subsection{X-ray Luminosities} 

In order to estimate the X-ray background on each image, we extracted
strips from each edge of the frame (excluding a 5-pixel border).  Each
of these strips was smoothed with the 250$\hratio$ kpc Gaussian
function. The four strips were then median-filtered using a
sigma-clipping algorithm to reject deviant pixels.  The median and
$1\sigma$ deviation were then determined for each frame from the
statistics of the coadded strips. We found an average background
count-rate of $1.95\pm0.50\times10^{-5}~\rm counts/sec/pixel$ (pixels
were 15$\arcsec \times$15$\arcsec$), or a mean surface brightness of
$3.12\times 10^{-4}~\rm counts~sec^{-1}~ arcmin^{-2}$.

For each X-ray peak, we calculated the detection threshold for each
annulus of the intensity profile (as well as the cumulative profile)
as the total background-subtracted counts from the source divided by
the square-root of the background counts (${C}\over{\sqrt B}$).  We
also computed a total $S/N$ including the Poisson error from the
source (added in quadrature), in order to estimate the error on the
integrated luminosity.  

Conversion from count-rate to flux was accomplished using energy
conversion factors (ECFs) derived for each cluster.  The ECFs were
determined using XSPEC, assuming an absorbed Raymond-Smith plasma with
0.3 solar abundance, a constant temperature of 5 keV, and convolved
with the response matrix of the PSPC-C detector.  The galactic $\rm
N_H$ column densities were taken from the {\sl HI} map of Lockman
\citep{lockman90,h1map}.  Because of the soft-energy response of the
PSPC these factors are not particularly sensitive to the assumed
temperature. The difference in ECFs for $T=3$ and $7~keV$ results in
less than a 3\% difference in the calculated flux.  As the clusters
are limited to a narrow range in galactic absorption the ECFs only
vary from $8.52-8.66\times 10^{-12}$, with an average value of
$8.61\pm0.03 \times 10^{-12}~\rm ergs~cm^{-2}~count^{-1}$. The X-ray
luminosity was then calculated in the usual way from the Luminosity
distance of the cluster.

\subsection{Extended and Point-Source Discrimination - Finding the Clusters} 

From the intensity profiles we have defined two different {\it
  flux-ratios} in order to parameterize the extent and concentration
of the X-ray emission for each peak.  Because the {\sl IRING} data are
binned into circular annuli of separation 31.25$\hratio$ kpc, we have
chosen a minimal aperture of 2 bins or 62.5$\hratio$ kpc.  The flux
contained within this aperture is compared to that enclosed within
apertures of both 250 and 500$\hratio$ kpc in size
($f_{250}={{F_{250}}\over{F_{62.5}}}$ and
$f_{500}={{F_{500}}\over{F_{62.5}}}$).

We empirically determined the expected flux-ratios for a point source
based on inspection of isolated X-ray peaks in a number of cluster
fields. In Figure 3 we plot $f_{250}$ versus $f_{500}$ and $f_{500}$
vs. redshift for these peaks. We verify that there is no dependence of
$f_{500}$ or $f_{250}$ with redshift for point sources.  While the
aperture size increases with redshift, as we chose isolated peaks, the
ratio is independent of aperture size as long as there is no
significant contamination. We conservatively set limits of $f_{500}
\geq 4$ and $f_{250} \geq 2$ in order to classify a peak as extended.

As a first pass in associating the catalogued peaks with the clusters
we applied a positional constraint such that the expected X-ray
centroid should be within $750\hratio$ kpc of Abell's catalogued
cluster position.  At the median redshift of the sample this
corresponds to 10.8 arcminutes, which is conservative enough to avoid
missing any clusters in the first round of processing.  After
producing a list of candidate X-ray sources which met both the
flux-ratio cutoffs for being extended and this positional constraint,
we visually inspected the optical/X-ray overlays (see Figures 2a,b,c)
to confirm a positional coincidence with a cluster.  From the final
tabulated list of X-ray cluster peaks, we find a median optical/X-ray
offset of $188\hratio$ kpc, which at the median redshift corresponds
to an offset of just 2.6 arcminutes.

We also measured for each cluster the X-ray luminosity in a
500$h_{75}^{-1}$ kpc aperture centered on Abell's optical position,
regardless of whether there was an actual X-ray peak at that location.
For clusters with no significant X-ray peaks (or only those associated
with point sources), upper-limits were calculated based on $3\sigma$
times the error in the background integrated over the 500$\hratio$ kpc
aperture.

\section{Results} 

The results of the X-ray analysis for each cluster is listed in Tables
1 and 2.  For Table 1, the first line for each cluster includes the
(1) cluster name, (2) redshift, (3) mean RASS exposure time in
seconds, (4,5) optical position (RA,DEC) of the cluster center
(\citep{aco}) in J2000 coordinates, (8) the X-ray luminosity and error
measured within $500\hratio$ kpc centered on Abell's position or an
upper-limit if no significant X-ray peaks were detected, and (10) the
total S/N within this aperture.  Subsequent lines for each cluster
list (4,5) the X-ray position (RA,DEC) of the peak(s) in J2000
coordinates (6) the projected positional offset from Abell's cluster
position in $\hratio$kpc, (7) the flux and error within a $500\hratio$
kpc aperture, (8) the X-ray luminosity and error calculated from the
flux, (9) the flux ratio ($f_{500}$) measured from the
surface-brightness profile (see definition above), (10) the
significance (S/N) of the detection integrated over the 500$\hratio$
kpc aperture, and (10) the maximum S/N within this aperture in any of
the 16 annular bins.  The order of these subsequent lines for each
cluster is first the most likely identification of the cluster
emission followed by any other candidates in order of increasing
distance from Abell's position.  In some cases there is an ambiguity
between multiple X-ray peaks of similar brightness.  We list all such
peaks in these cases.  All peaks listed in Table 1 meet the criteria
discussed in the previous section regarding offsets from Abell's
positions and the measured flux-ratios, which insures that the peaks
are resolved and possible candidates for the cluster emission.

In Table 2 we list all significant X-ray peaks found in the direction
of all clusters within the $2\times 2\hratio$ Mpc search box.  In
columns 1-3 we list the Abell number of the field and RA, DEC of the
X-ray peak.  In the remaining columns we list (4) the offset in arcsec
between the X-ray peak and the optical position of the cluster center,
(5) the peak flux in $counts/sec/pixel$, (6) the flux-ratio $f_{250}$
(see definition in the previous section), (7) the significance of the
detection ($({{C}\over{\sqrt{B}}})_{250}^{max}$), and (8) a code if
applicable identifying the peak with the cluster emission (C).
Multiple peaks are labeled (M) if the cluster emission appears bimodal
or multimodal.

Based on these classification criteria, we find a cluster detection
rate of 83\%.  In Figure 4 we show a histogram of the X-ray
luminosities and separately, the upper limits.  We find an average and
median $L_X$ of $2.04\pm 0.22$ and $0.99\pm 0.23$ $\lhratio 10^{43}$
ergs sec$^{-1}$, respectively.  The precipitous decline in the number
of clusters for $L_X < \lhratio 10^{43}$ ergs sec$^{-1}$ was shown
\citep{led99} to be a result of incompleteness due to the optical
selection of our sample (our sample is not X-ray flux-limited).  While
Abell's catalog is complete in an optical sense over these redshifts,
there are a large number of optically poorer clusters with X-ray
luminosities in the range of Richness class=0 Abell clusters.

We compare our measured fluxes to those from previous work in Figure
5.  We have cross-correlated our sample with the XBACS \citep{xbacs}
and BCS+eBCS \citep{bcs,ebcs}.  A number of conversions are needed in
order to directly compare the fluxes.  These other samples report
fluxes over the energy band (0.1-2.4 keV) and have applied statistical
corrections for missing low-surface-brightness emission and for
point-source contamination. For the comparison we adjust our fluxes
(measured within a $500\hratio$ kpc radius aperture to a total flux
following \cite{briel93} and as described in \cite{led99} assuming a
core radius of 250 kpc for all clusters.  This results in a count-rate
correction factor of $\sim 1.8$, very close to that derived for the
BCS (see fig 11 in \cite{bcs}) with a similar source extent (with a
large scatter). We convert the BCS+eBCS and XBACS fluxes to 0.5-2.04
keV using the individual cluster temperatures given in their tables.
In general the agreement is quite good; the dispersion increases with
increasing flux as one would expect if the scatter is primarily due to
differences in the correction for missing flux.

\subsection{Flux-Ratios and Cooling Flows} 

In Figure 6, we plot both $f_{500}$ (solid-histogram) and $f_{250}$
(dashed-histogram) for all detected clusters using the selected
cluster peak as listed in Tables 1 and 2.  For the entire sample, we
find $<f_{500}> = 12.1\pm6.9$ and $<f_{250}> = 6.3\pm3.4$.  Thus, the
vast majority of the clusters are significantly extended as compared
to a point-source model ($f_{500} < 4$, $f_{250} < 2.0$).

As one of the observational signatures of a cooling flow is a central
X-ray excess, these flux-ratios may be useful in finding clusters with
compact cores. We have cross-correlated our sample with the
Cooling-Flow cluster lists of \cite{arnaud}, \cite{jf99},
\cite{dwhite97}, and \cite{peres}.  From these lists, we include a
cluster as a cooling-flow if the derived $\dot M$ was $> 0$ and the
cooling time was less than a Hubble time, as tabulated in these
papers.  For clusters with both {\sl Einstein} and {\sl ROSAT}
measurements, we preferentially choose the {\sl ROSAT} numbers.
Similarly, we adopt the values determined from the {\sl ROSAT} PSPC
over the HRI if both are available.  In Figure 7, we plot histograms
of $f_{500}$ for (a) clusters with cooling-flows, (b) no cooling-flow
clusters, and (c) all other detected clusters in our sample for which
no deprojections have been performed.  Those clusters classified as
{\it no cooling-flow} are labeled as such because the deprojection
analysis of these clusters indicates a cooling time greater than a
Hubble time. We find that $f_{500}$ is biased towards smaller values
for the cooling-flow clusters. The 75th, 50th, and 25th percentile
values are for Cooling Flow Clusters: 5.8, 8.6, and 12.4,
respectively, with a mean of 9.7$\pm$0.7.  For Non Cooling Flows
these are 9.3, 12.9, and 19.6 with a mean value of 15.3$\pm$1.2.  We
also ran several two-sample tests to compare the $f_{500}$
distributions (Log-Rank, Gehan Wilcoxon, KS, etc.). All tests
consistently find the two samples different at the $>$99\% level.  We
find no correlation between $\dot M$ and $f_{500}$.

We plot $f_{500}$ vs. redshift in Figure 8 with different symbols to
indicate cooling flows (C), non cooling flows (N), and other clusters
(small O). No significant correlations are found.  The effective
resolution in the RASS is $\approx 45\arcsec$.  The smaller 62.5kpc
aperture used in our flux ratios corresponds to $42\arcsec$ at z=0.09
and hence is a good match to the PSF. We therefore don't expect
distance-related biases in our analysis. 

While the $f_{500}$ distributions for clusters with and without
cooling flows are statistically different, the range of values overlap
significantly making it difficult to use $f_{500}$ alone as an
indicator for a cooling flow.

\subsection{Detection Statistics and Optical Cluster Properties}

For the full sample we find an X-ray detection rate of 83\%.  As a
function of Abell Richness Class we find detection rates of:
RC0=76.6\% (105/137), RC1=89.6\% (111/124), RC2=91.6\% (22/24), and
RC3=100\% (1/1).  The higher detection rates for richer clusters are
consistent with increasing gas mass and a deeper gravitational
potential in richer systems.  Thus, we might expect to find a
correlation between $L_X$ and cluster richness.  This relationship has
been examined by \cite{dfj99}(DFJ99), \cite{briel93}, \cite{johnson},
and \cite{bahcall77} among others, with varying results depending on
the sample size and selection. Most, but not all of these previous
studies preferentially selected the most X-ray luminous or richest
clusters. Most recently, \cite{yee2003} show that the richness
(parameterized by B$_{gc}$; the galaxy cluster center correlation
amplitude) is a key defining characteristic, tightly correlated with
the velocity dispersion and X-ray temperature.  We expect that the
observed scatter between $L_x$ and richness is directly related to the
known scatter between $L_x$ and temperature, and more fundamentally
on the cluster mass.

In Figure 9 we show the relationship between $L_X$ and Abell's
richness for our sample.  Statistical tests on these data indicate a
positive correlation.  Both the Spearman-$\rho$ (rank-correlation) and
Kendall-$\tau$ statistics derive a probability of $< 1\times 10^{-4}$
that $L_X$ and Richness are uncorrelated ($\rho=0.544, \tau=0.715$).
We also show in Figure 9 the results of fitting the data.  The
solid-line is the result when the upper-limits are not used and the
data points are weighted inversely by their standard deviations
($L_{X}(h_{75}^{-2}10^{43}) = 0.002\pm0.0002 \times R^{1.77\pm0.01}$).
The dashed-line is the result of a linear-regression fit
(Buckley-James method)) which includes the censored data points
($L_{X}(h_{75}^{-2}10^{43}) = 0.0007^{+0.0011}_{-0.0004} \times
R^{1.79\pm0.19}$).  The two fits agree within the estimated errors.
Our derived slope is nearly identical to that of \cite{briel93} when
the difference in $H_0$ is taken into account.

There are wide-spread disagreements amongst previous studies as to the
true correlation between $L_X$ and cluster morphology, as it is
difficult to disentangle the dominant factors given that morphology
and richness are also correlated (at least weakly).  Conclusions range
from the cluster morphology having equal importance (DFJ99;
\cite{bahcall77}) in determining $L_X$ to only a very weak dependence
\citep{johnson,jf78}.  In common to these studies, however, is the
finding that the early-type cluster morphologies (BM type {\sl I,I-II} and
RS type {\sl cD,B}) may have an independent influence on the luminosity,
presumably due to the presence of a cD-like galaxy and correspondingly
deeper gravitational potential \citep{mch78}.  McHardy found, however,
that cD-like galaxies in poorer clusters (such as AWM \citep{awm}, MKW
\citep{mkw}, and WP \citep{wp} clusters) have a much lower probability
of X-ray emission as compared to Bautz-Morgan type {\sl I} Abell
clusters.  Thus, the $L_X$/richness dependence would appear to be
stronger based on these arguments.  To complicate matters further,
DFJ99 found that the $L_X$-richness correlation primarily pertains to
{\bf late} BM type clusters, and $L_X$-richness are in fact
uncorrelated in early BM type clusters.  Their study, however, was
limited to the richest (RC$\geq$2) Abell clusters. 

We have looked for correlations in the detection statistics of
$L_X$ with optical cluster morphology quantified by the RS and BM
classifications.  We show these results in Table 3.  First, we have
examined the $L_X$-richness relationship for the BM and RS cluster
types individually.  We find that each of the individual classes has a
significant $L_X$-richness correlation (at $>$ than 99.99\%
probability), surprisingly with the exception of the BM {\sl I-II} and
RS {\sl B}-type clusters.  For these two classes, the probabilities
are 13\% and 6\% based on Spearman-$\rho$ and 23\% and 4\% from
Kendall-$\tau$ for BM and RS types respectively, that $L_X$ and
richness are uncorrelated.  However, it is unclear if this is actually
significant since both of these classes have 100\% detection
rates in our sample and also have the lowest number statistics of any
class.  The Spearman-$\rho$ test in particular is known to return
questionable results for $N<30$ data points \citep{asurv}.  When we
divide the sample into early/late-type bins (early type = RS {\sl
  cD,B,C,L} and BM {\sl I,I-II}; DFJ99) we find that $L_X$ and
richness are correlated at $\approx$ equal significance for both early
and late-type clusters, in contrast to the results of DFJ99.  Recall,
the DFJ99 sample only includes RC$\geq$2 clusters while our sample
includes all richness classes.

In the third and fourth columns of Table 3 we list the average $L_X$
and richness of each of the morphological classes.  From simple
inspection, it is clear that early cluster types appear to have higher
$L_X$ than late-types by a fairly large factor.  There is a trend for
the late-type clusters to have lower richness. To quantify these
results in a more robust way, we have applied a number of two-sample
statistical tests in order to estimate the probabilities that $L_X$
and richness for the different BM and RS classes are derived from the
same parent sample.  In this way, we can separate the possible mutual
dependence of $L_X$ on richness and morphology to determine which is
more important.

For cluster richness, we find marginal evidence that the
early/late-type BM or RS clusters are culled from different parent
populations (probabilities $<$ 5\% for BM and $<$ 2\% for RS), while
for $L_X$ we find that the probability of early/late-type clusters
having the same $L_X$ distribution is $< 1-2\times 10^{-4}$.  Thus,
the statistical results suggest that the cluster morphology is in fact
{\bf more} important than the richness in determing $L_X$ for a given
cluster, contrary to some other findings (DFJ99 suggest at least equal
importance).  As a consistency check, assuming $L_X \propto R^{1.7}$,
the difference in richness over the range of cluster types would
predict a factor of 1.3 increase in $L_X$ from late to early BM or RS
types, while we see an increase of $\approx$ a factor of 4.

The presence of a cooling flow can also perturb the scaling relations
for clusters.  \cite{fabian94} showed that clusters with cooling flows
have either a lower temperature or a higher luminosity than those
without; the offset from the mean $L_x/T_x$ scaling relation being
proportional to $\dot M$. As cooling flow clusters are thought to tend
towards early type morphologies \citep{edge91} and the fraction of
cooling flow clusters is measured to be $\geq 70\%$ \citep{peres}, the
effects we are seeing may be directly related to cooling flows.  The
percentage increase in $L_X$ resulting from a cooling flow is expected
to be of order $<30\%$, similar to the factor of 1.3 we expect from
differences in richness from late to early type clusters. In Figure 10
we show the distribution of RS and BM types for cooling flow and
non-cooling flow clusters. While based on similar criteria, it is
clear that the RS and BM classifications do not exactly parallel one
another for the same clusters.  With regard to RS type, 76\% of the
cooling flow clusters are in early-type (cD (50\%) or C) clusters. The
cooling flow/non cooling flow RS distributions are different at the
97\% level.  Surprisingly, however, 39\% of the cooling flows are in
BM type III clusters (corresonding to mostly C,F RS types).  The BM
distributions are not statistically different between the two samples.

Because the optical morphological classifications are based on the
spatial distribution of the brightest galaxies, the effects of
fore-/back-ground contamination can produce a bias with redshift. We
have checked the redshift distributions of the RS classes separately
and grouped into the early/late-type categories given above. A KS-test
shows that the two categories are consistent with having been drawn
from the same redshift distribution.

Based on our results from the X-ray properties, we would argue for a
modification to the traditional binning of early and late type
clusters.  For the BM classes, BM types {\sl I, I-II}, and {\sl II}
have very similar X-ray properties, and the BM {\sl II-III} and {\sl
  III} classes have nearly identical (and lower) $L_X$ and X-ray
detection rates.  For the RS classes, the {\sl cD} and {\sl B} types
clearly stand out as having higher $L_X$, whereas the {\sl C, L, F},
and {\sl I} group all have lower but essentially the same $L_X$.  As
the X-ray emission is a better tracer of the underlying gravitational
potential and cluster mass, we would argue for this binning scheme
over previous, optically based early/late type bins.

The interpretation of these results must be related to both the
presence of cooling flows and to the expected evolution in $L_x$ as
the cluster becomes more and more centrally concentrated over time and
with successive mergers \citep{bryan98}.  Approximately 40\% of all
Abell clusters show evidence of recent mergers based on X-ray and
optical substructure \cite{jf99}, \cite{kolokotronis}, and
\cite{mcnamara}.  Cluster merger simulations predict significant
changes in $L_X$ as the gravitational potential is initially weakened
in strength early on and later is enhanced over the pre-merger value
after a few crossing times \citep{ricker,kurt97,kurt96}.  These
predictions are therefore consistent with our findings, assuming that
{\it early}-type clusters are in fact the most evolved systems and
{\it late}-type systems the least.

\section{Conclusions} 

We report an X-ray analysis of a complete sample of 288 nearby ($z\leq
0.09$) Abell clusters from the {\sl ROSAT} All-Sky Survey.  We find a
detection rate of 83\% at greater than 3$\sigma$ confidence and
tabulate fluxes and luminosities, or upper limits for the entire
sample.  For each cluster we have identified the most likely X-ray
peak and position which should be identified with the cluster
emission.  We have checked X-ray/optical overlays with the Palomar
Digitized Sky Survey to attempt to confirm the association with the
clusters.  All X-ray peaks within a $2\times 2 \hratio$ Mpc box
centered on Abell's position have also been catalogued. Additionally,
we have made optical/X-ray overlays of all cluster fields available
over the WWW at \url{http://put.the.rightaddress.here}.

We have looked for correlations between the X-ray and optical cluster
properties.  We find significant correlations between the cluster
richness, optical morphology and $L_x$. Early cluster types (BM types
I,I-II and RS types cD, B) have systematically higher X-ray
luminosities than later types. Our results suggest a slight revision
of the early/late-type morphological separation into these bins. The
results can be explained by the large fraction of cooling flows in
clusters ($\geq 70\%$) and a preference for clusters with early-type
optical morphology.

The measured X-ray flux ratio ($f_{500}$) for clusters with and
without cooling flows are statistically different. Coupled with an
early RS morphological type, $f_{500}$ is potentially useful in
identifying cooling flow candidates,

Papers already published using this sample include an analysis of the
X-ray luminosity function and comparison to cosmological models
\citep{led99}, and a computation of the two-point spatial correlation
function \citep{miller99}.

\acknowledgements

We thank the referee, Alastair Edge, for his comments which greatly improved
the presentation in this paper.  This work was supported in part by
NASA grants NAG5-6739, NAG5-10516 to ML, NAGW-3152 to JB and FO, and
NSF grants AST-9317596 and AST-9896039 to JB The ROSAT Project is
supported by the Bundesministerium f\"ur Bildung und Forschung
(BMBF/DLR) and the Max-Planck-Gesellschaft (MPG).  This research was
supported by the Gemini Observatory, which is operated by the
Association of Universities for Research in Astronomy, Inc., on behalf
of the international Gemini parthership of Argentina, Australia,
Brazil, Canada, Chile, the United Kingdom and the United States of
America.  We acknowledge use of the NASA/IPAC Extragalactic Database
(NED), which is operated by the Jet Propulsion Laboratory, California
Institute of Technology, under contract with NASA.

\clearpage

\centerline{\bf FIGURE CAPTIONS} 

\figcaption[]{(A) The derived selection function for our Abell cluster
  sample.  The aitoff grid is show in Galactic coordinates.  The
  solid-line is the $\delta=-27^{\circ}$ cutoff of Abell's northern
  catalog. We estimate a sky-coverage of 34\% for our sample. (B) The
  distribution of all Abell clusters in the sample on the same grid
  for comparison}

\figcaption[]{X-ray/optical overlays from a representative group of 6
  clusters from the sample with average $L_X$.  We will make these
  overlays for all clusters available over the WWW. In Fig. 2a we show
  overlays for A193 and A602 (top-bottom), Fig. 2b: A1035 and A1541,
  and Fig 2c: A1650 and A2124.}

\figcaption[]{We plot the flux-ratios $f_{250}$ vs. $f_{500}$ for a
  sample of isolated pointsources. In the lower panel is plotted
  $f_{500}$ vs. redshift. We conservatively set $f_{250}\leq 2$ and
  $f_{500}\leq 4$ as the range of values consistent with an unresolved
  peak.}

\figcaption[]{Histogram of the X-ray luminosity of our cluster
  detections (solid) and upper-limits (dashed).}

\figcaption[]{Comparison of total cluster X-ray flux from the BCS,
  eBCS, and XBACS surveys to those in this paper.  See the text for an
  explanation of the correction factors (and uncertainties) applied to
  compare the fluxes on a similar scale.}

\figcaption[]{Histogram of the flux-ratios for all cluster X-ray
  detections.  The solid-line is for $f_{500}$ as defined in the text.
  The dashed line is for $f_{250}$.}

\figcaption[]{(a) Histogram of $f_{500}$ for cooling flow clusters,
  (b) clusters without cooling flows, and (c) clusters from our sample
  with no deprojection analysis in the literature.}

\figcaption[]{$f_{500}$ vs redshift.  The different symbols are
  cooling flow clusters (C), non cooling flow clusters (N), and all
  other cluster detections.}
  
\figcaption[]{$L_X$ plotted versus Abell richness.  Filled circles are
  detections, the open-symbols are upper-limits.  The solid-line is a
  weighted least-squares fit to the data (excluding the
  non-detections).  The dashed-line is a linear-regression fit which
  includes the censored data points. At the top of the plot, the range
  for Abell Richness classes 0-3 are given.}

\figcaption[]{Histograms of Rood Sastry (RS; solid line) and
  Bautz-Morgan (BM; dashed line) morphological types for cooling flow
  and non cooling flow clusters.}

\end{document}